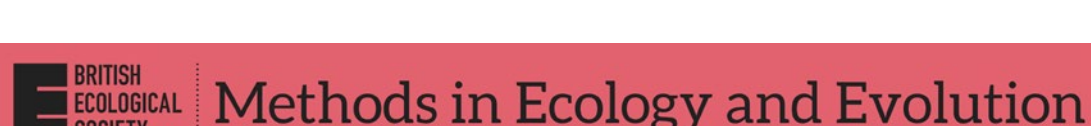

RESEARCH ARTICLE

# animal2vec and MeerKAT: A self-supervised transformer for rare-event raw audio input and a large-scale reference dataset for bioacoustics

**Julian C. Schäfer-Zimmermann**[1,2,3] | **Vlad Demartsev**[1,2,3,4] | **Baptiste Averly**[1,2,3,4] | **Kiran L. Dhanjal-Adams**[1,2,3,5] | **Mathieu Duteil**[1,2,3] | **Gabriella Gall**[1,2,3,6] | **Marius Faiß**[1,2] | **Lily Johnson-Ulrich**[4,7] | **Dan Stowell**[8,9] | **Marta B. Manser**[4,7,10] | **Marie A. Roch**[11] | **Ariana Strandburg-Peshkin**[1,2,3,4]

[1]Department for the Ecology of Animal Societies, Max Planck Institute of Animal Behavior, Konstanz, Germany; [2]Department of Biology, University of Konstanz, Konstanz, Germany; [3]Centre for the Advanced Study of Collective Behaviour, University of Konstanz, Konstanz, Germany; [4]Kalahari Research Centre, Van Zylsrus, Northern Cape, South Africa; [5]Royal Botanic Gardens Kew, Richmond, UK; [6]Zukunftskolleg, University of Konstanz, Konstanz, Germany; [7]Department of Evolutionary Biology and Environmental Studies, University of Zurich, Zurich, Switzerland; [8]Department of Cognitive Science and Artificial Intelligence, Tilburg University, Tilburg, The Netherlands; [9]Naturalis Biodiversity Center, Leiden, The Netherlands; [10]Interdisciplinary Center for the Evolution of Language, University of Zurich, Zurich, Switzerland and [11]Department of Computer Science, San Diego State University, San Diego, California, USA

**Correspondence**
Julian C. Schäfer-Zimmermann
Email: jzimmermann@ab.mpg.de

**Funding information**
Minerva Foundation; Universität Zürich; MAVA Foundation; H2020 Marie Skłodowska-Curie Actions (BioacAI), Grant/Award Number: 101071532; Max-Planck-Gesellschaft; Human Frontier Science Program, Grant/Award Number: RGP0051/2019; Alexander von Humboldt-Stiftung; Young Scholars Fund at the University of Konstanz; Natural Environment Research Council, Grant/Award Number: NE/G006822/1; Centre for the Advanced Study of Collective Behaviour, Grant/Award Number: EXC 2117-422037984; H2020 European Research Council, Grant/Award Number: 294494 and 742808; Gips-Schüle-Stiftung

**Handling Editor:** Aaron Ellison

## Abstract

1. Bioacoustic research, vital for promoting conservation and understanding animal behaviour and ecology, faces a monumental challenge: analysing vast datasets where animal vocalizations are rare. While deep learning techniques are becoming standard, adapting them to bioacoustics remains difficult.
2. We address this challenge with animal2vec, an interpretable large transformer model and a self-supervised training scheme tailored for sparse and unbalanced bioacoustic data. It learns from unlabelled audio and then refines its understanding with labelled data. Furthermore, we introduce and publicly release MeerKAT: **Meer**kat **K**alahari **A**udio **T**ranscripts, a dataset of meerkat (*Suricata suricatta*) vocalizations with millisecond-resolution annotations, the largest labelled dataset on a non-human terrestrial mammal currently available.
3. Our model sets a baseline on the MeerKAT corpus, outperforming other transformer models, and improves on existing methods on the publicly available NIPS4Bplus birdsong dataset. Moreover, animal2vec performs well even with limited labelled data (few-shot learning).
4. animal2vec and MeerKAT provide a new reference point for bioacoustic research, enabling scientists to analyse large amounts of data even with scarce ground truth information.

Marie A. Roch and Ariana Strandburg-Peshkin shared senior authorship.

KEYWORDS
animal vocalizations, bioacoustics, reference dataset, self-supervised deep learning, transformer

## 1 | INTRODUCTION

Bioacoustics, the study of animal sounds, reveals invaluable insights into the behaviour (Bradbury & Vehrencamp, 1998; Demartsev et al., 2023; Rutz et al., 2023) and ecology (Fleishman et al., 2023; Penar et al., 2020; Rasmussen et al., 2024) of animal species, with important implications for conservation (Laiolo, 2010; Mcloughlin et al., 2019). Automated analysis of acoustic recordings can greatly advance the types of questions that can be asked by enabling annotation of long-duration recordings. Despite the broad potential of bioacoustic datasets, events of interest such as vocalizations are often sparse, brief and in noisy conditions, making manual as well as automated analysis challenging (Allen et al., 2021; Lindseth & Lobel, 2018; Loo et al., 2025; Lostanlen et al., 2018; Madhusudhana et al., 2022; Morfi et al., 2019; Ness et al., 2013; Sugai et al., 2019; Wall et al., 2021).

Deep learning is a common approach to tackle large and densely labelled datasets (LeCun et al., 2015), and recently, transformer-based models (Vaswani et al., 2017) have achieved state-of-the-art results across many tasks and modalities (Lin et al., 2022). However, there is a lack of such large-scale datasets and training approaches for sparse data using next-generation transformer-based models within bioacoustics (Stowell, 2022).

Currently, in bioacoustics, the primary data (audio waveforms) are usually feature-engineered into spectrograms for input to convolutional neural network models (CNNs) originally designed for computer vision (Stowell, 2022). However, using spectrograms and CNNs is justified more by empirical success than conceptual fitness. Spectrograms challenge the notion of translational invariance in CNNs (Wyse, 2017), discard phase information or temporal fine structure (Stowell, 2022), and the commonly used Mel-scale biases the input towards human hearing (Stowell, 2022). Further, in computer vision, attention-based encoder-only visual transformers (ViTs) (Dosovitskiy et al., 2020; Khan et al., 2021) have replaced CNNs, excelling through large-scale pretraining on densely labelled datasets. Pretraining is a method of learning a general model, which can then be fine-tuned on downstream tasks. This training paradigm is referred to as pretrain/fine-tune, where the gold standard is supervised pretraining (Dosovitskiy et al., 2020; Radford, 2023). Supervised pretraining requires large, diverse and fully labelled datasets (e.g. Imagenet for computer vision (Deng et al., 2009)). However, this strategy is not feasible in bioacoustics due to limited labelled dataset size. The largest publicly available labelled bioacoustic dataset is the iNaturalist sounds dataset (Chasmai et al., 2024), with 1200 h across 230,000 samples and Birdset (Rauch et al., 2024), with 6800 h across 530,000 samples. However, both datasets are weakly- or only roughly labelled (Zhou, 2018) (no, or only very coarse onset/offset markers), and are predominantly based on community science uploads that only partially reflect realistic bioacoustic recording scenarios.

Self-supervised learning can provide an alternative to supervised pretraining (Chen et al., 2022; Liu et al., 2021; Longpre et al., 2023), where the generalist model is trained using an artificial supervisory task created from the data without using any ground truth labels (Liu et al., 2021).

Currently, contrastive-learning-based (CLR; learning from differences between examples) pretraining is the dominant scheme in computer vision (Chen, Kornblith, Norouzi, & Hinton, 2020; Chen, Kornblith, Swersky, et al., 2020; He et al., 2020; Robinson, 2021; Tomasev et al., 2022) and audio processing (Baevski et al., 2020; Niizumi et al., 2021, 2023; Saeed et al., 2021), whereas generative methods (learning by reconstructing), either autoregressive (Brown et al., 2020) or bidirectional mask-prediction (Devlin et al., 2019; Hsu et al., 2021; Liu et al., 2019), yield state-of-the-art results in natural language processing. However, these approaches are conceptually ill-equipped to handle sparse and unbalanced bioacoustic data. Generative pretraining is known to diverge when faced with sparse and noisy data (Lin et al., 2021; Zhu et al., 2022) and CLR-based methods suffer from so-called *easy negative sampling* (Chung et al., 2021; Jaiswal et al., 2020), where a model struggles to converge as the small number of relevant signals is too easy to identify compared to the irrelevant bulk of the data, which, in return, leads to little contribution to the contrastive loss function from the relevant signals.

Despite these obstacles, ViTs have recently been introduced to bioacoustics (Gong et al., 2021; Gu et al., 2024; Hagiwara, 2023; Rauch et al., 2025; Robinson, Miron, et al., 2024; Robinson, Robinson, & Akrapongpisak, 2024; Wolters et al., 2021; Wyatt et al., 2021; You et al., 2023), where approaches range from no pretraining (Wolters et al., 2021; Wyatt et al., 2021) to various pretraining strategies, including web-scraped human speech (Gu et al., 2024), human language audio-caption pairs (Robinson, Miron, et al., 2024; Robinson, Robinson, & Akrapongpisak, 2024), pretraining on ImageNet (Deng et al., 2009; Gong et al., 2021), repurposing pretrained CNNs from Audioset (Gemmeke et al., 2017) as a pre-transformer feature extraction step (You et al., 2023) or training smaller architectures using smaller bioacoustic datasets (Hagiwara, 2023). However, as of now, pretraining a large transformer model with bioacoustic data itself remains an open problem.

In sum, deep learning in bioacoustics faces multiple challenges: First, the inherent limitations of spectrographic representations; second, the lack of large-scale fully labelled datasets for supervised pretraining; and third, the conceptual problems of prevailing self-supervised pretraining strategies, such as CLR, with sparse, noisy and unbalanced bioacoustic data.

We address these challenges by releasing the animal2vec framework and the MeerKAT dataset.

animal2vec is a framework for training animal call recognizers from raw waveforms containing sparsely distributed calls with non-uniformly distributed call types. It is conceptually simple, built for noisy and sparse datasets, achieves state-of-the-art performance, is capable of learning from limited labelled training data (few-shot learning) and provides temporal and spectral interpretability.

MeerKAT: **Meer**kat **K**alahari **A**udio **T**ranscripts is a 1068h large-scale dataset, of which 184h are strongly labelled, that exhibits realistic sparsity conditions containing data from audio-recording collars worn by free-ranging meerkats (*Suricata suricatta*) and boom-mic recordings at the Kalahari Research Centre, South Africa (The Kalahari Research Centre (KRC), n.d.), making it the largest publicly available strongly labelled bioacoustic dataset on non-human terrestrial mammals to date.

Here, we first describe the features of the MeerKAT bioacoustic dataset and the animal2vec framework in subsections 2.1 and 2.2 in the Section 2.

We then evaluate the performance of animal2vec on the MeerKAT dataset against two transformer-based approaches (Baevski et al., 2023; Gu et al., 2024) and the publicly available strongly labelled set of bird calls, NIPS4Bplus, comparing it with published results (Bravo Sanchez et al., 2021; Morfi et al., 2019) in the Section 3. Furthermore, a number of additional experiments are performed. Due to space considerations, we report these in a designated section of additional experiments (Supporting Information: Section 'Additional experiments') in the supplemental information and only briefly mention them in the main text.

We release all the code, model weights and data as open access (see data availability statement). Furthermore, we provide a non-technical introduction for readers without a machine learning background in Supporting Information: Section 'Detecting and classifying animal calls from audio data using animal2vec'.

Our work (i) paves the way to adapt and specialize next-generation transformer models to the domain of bioacoustics using the unified animal2vec framework, (ii) allows researchers with limited labelled data to classify large amounts of challenging data and (iii) introduces the first bioacoustic benchmark to evaluate large-scale pretrain/fine-tune approaches under realistic sparsity and class balancing conditions.

## 2 | MATERIALS AND METHODS

### 2.1 | The MeerKAT bioacoustic dataset

The compilation of MeerKAT reflects an extensive collaborative effort by researchers and students (see Acknowledgements) who recorded, labelled, and validated the dataset over an extended period. The data were collected during two field seasons (August–September 2017 and July–August 2019) at the Kalahari Research Centre (KRC) in South Africa. All procedures were approved by ethical committees of the University of Pretoria, South Africa (permit: EC031-17) and the Northern Cape Department of Environment and Nature Conservation (permit: FAUNA 1020/2016).

Meerkats are a social mongoose species native to the arid parts of southern Africa. Meerkats forage throughout the day by digging in the ground for prey, remaining cohesive with their group mates while moving within their territory. They use vocalizations to mediate a variety of social behaviours, and their vocal repertoire has been extensively characterized through decades of field research (Manser, 1998; Manser et al., 2014).

MeerKAT is released as 384,592 10-s samples, amounting to 1068h, where 66,398 10-s samples (184h) are labelled and ground-truth-complete; all call and recurring anthropogenic events in this 184h are labelled. All samples have been standardized to a sample rate of 8kHz with 16-bit quantization, which is sufficient to capture the majority of meerkat vocalization frequencies (the first two formants are below the Nyquist frequency of 4kHz; Townsend et al., 2014). The total dataset size of 59GB (61GB, including the label files) is comparatively small, making MeerKAT easily accessible and portable despite its extensive length. By agreement with the KRC, we have made these data available in a way that can further machine learning research without compromising the ability of the KRC to continue conducting valuable ecological research. Consequently, the filenames of the 10-s samples have been randomized, and their temporal order and individual identity cannot be recovered. However, this information can be requested from us.

In total, eight *vocalization* classes and three *miscellaneous* classes were identified. The vocalization classes are: *close call* (Townsend et al., 2010), *short-note call* (Collier et al., 2017; Demartsev et al., 2018), *social call* (Manser, 1998), *alarm call* (Manser, 2001), *aggressive call* (Collier et al., 2017), *move call* (Bousquet et al., 2011; Collier et al., 2017), *lead call* (Bousquet et al., 2011) and *other call* (see also (Manser, 1998; Manser et al., 2014) for a general overview on meerkat vocalizations). Meerkats can produce some calls that do not fit well into the set of described calls. These calls are frequently hybrid calls that bear similarity to multiple call types or are simply too rare to have their own category. Such calls are labelled as *other call* within MeerKAT. The three miscellaneous classes are for non-call events. The *synch* and *beep* events are generated by a GPS clock that was used to synchronize acoustic streams to one another across animals for the purposes of the behavioural study for which the data were collected (see 2d in Demartsev et al., 2024). The eating label indicates chewing noises from a successful foraging event. Figure 1 provides example spectrograms for a continuous audio stream and for every class. In addition to the vocalization and miscellaneous classes, a superordinate class called *focal* is used to indicate when a call was produced by the focal animal wearing the collar (or was the followed one in the case of boom-mic recording) as opposed to a nearby conspecific. Trained analysts made this decision based on the relative intensity of calls, changes in the frequency spectrum, and contextual information (see also supplemental information in Demartsev et al., 2024).

MeerKAT is *multi-class* and *multi-label*, which means that ground-truth labels may overlap. Labels are based on multiple annotators and have an inter-annotator reliability of a little over 10ms, which is consistent with inter-annotator reliability found in other datasets (Hallgren, 2012). All classes, as well as the distribution of the call durations, are shown in Figure 2.

**(a)** A realistic example

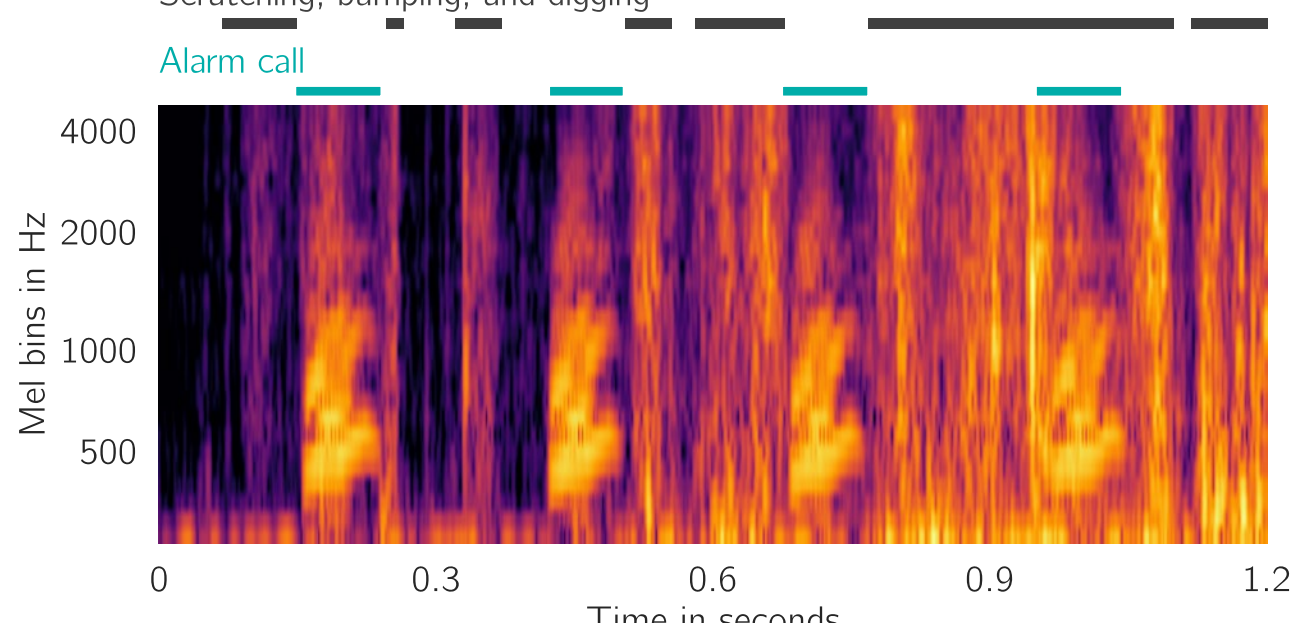

**(b)** Class examples with low noise

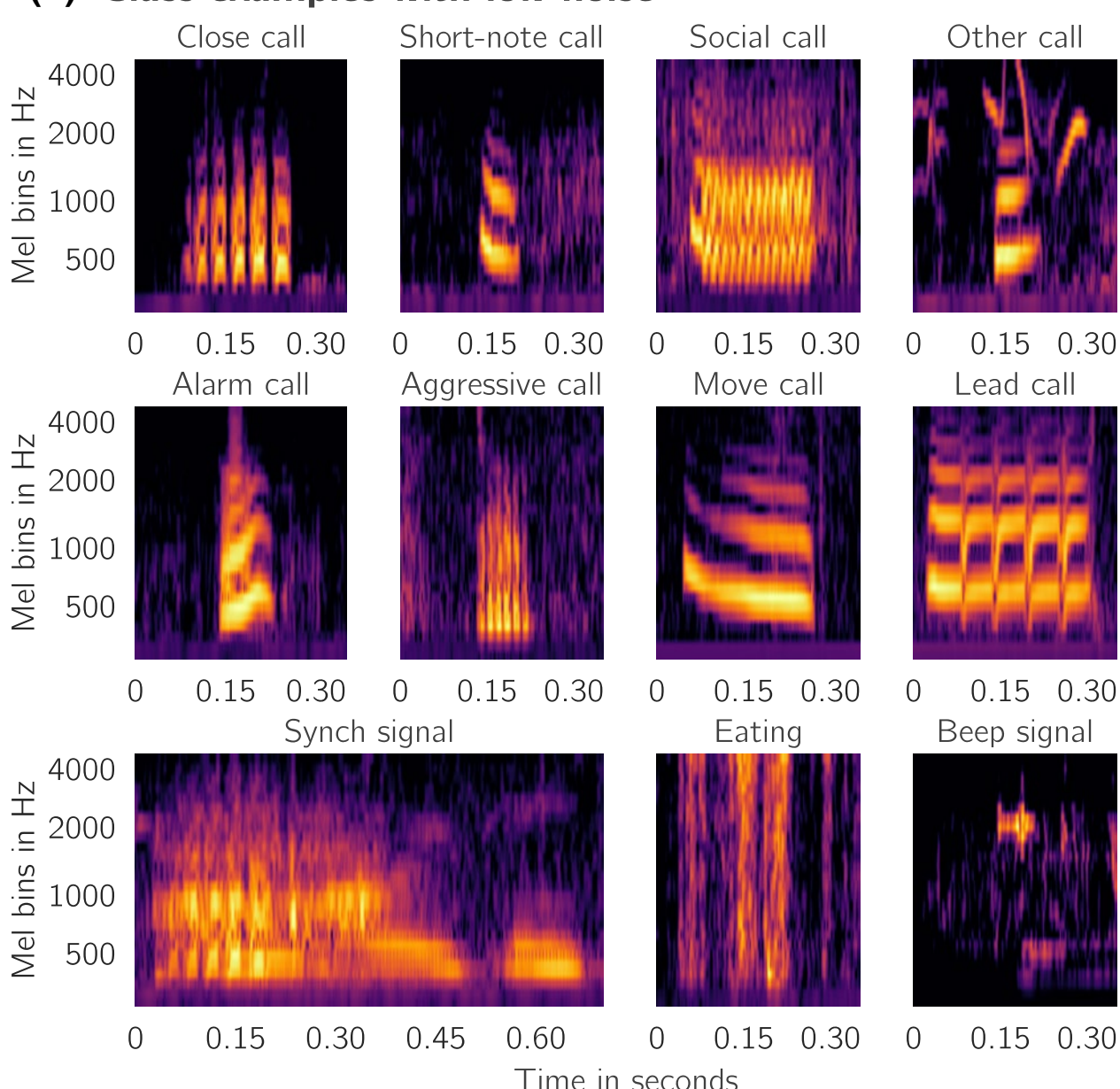

FIGURE 1 Example Mel spectrograms in dBr scale, where (a) is a representative stream of audio and (b) are the individual classes in MeerKAT. (a) It shows four alarm call events covered by a varying amount of spectrally broad, ultra-short and non-stationary noise patterns originating from the meerkats foraging for food by digging in the ground or bumping their collars into obstacles. Noise patterns such as these permeate the majority of MeerKAT. (b) It shows the spectral variability between classes, where the examples shown do not represent the overall data quality but reflect clean candidates.

Differences in the duration and frequency of events make analysing class abundance challenging. A class can be under-represented in terms of the number of examples while being simultaneously over-represented in terms of overall duration or vice versa. In the case of MeerKAT, both types of class imbalance are present, and moreover, they do not align. Therefore, there is no clear path to implement existing approaches to handle imbalance in this type of dataset (Chawla et al., 2002; He & Garcia, 2009).

Ultimately, MeerKAT makes a challenging benchmark for exploring sparsity, noise resistance, and imbalance in bioacoustics, containing events that are rare or plentiful, long or short, artificial or natural, temporally and/or occurrence-wise sparse, and spectrally rich, all while being covered in a large amount of spectrally broad, ultra-short, and non-stationary noise patterns (Figure 1).

The overall structure of MeerKAT, having a huge unlabelled and a large fully labelled subset originating from the same pool of audio files, makes it an ideal test-bed for pretrain/fine-tune approaches in bioacoustics. Additional information about recording, labeling, and pre-processing the MeerKAT dataset can be found in Supporting Information: Section 'MeerKAT audio and labels'.

## 2.2 | The animal2vec framework

### 2.2.1 | Design and pretraining scheme of animal2vec

animal2vec uses a mean teacher self-distillation framework (Baevski et al., 2022, 2023; Caron, 2021; Grill et al., 2020; Tarvainen & Valpola, 2017) similar to data2vec 2.0 (Baevski et al., 2023), which is known to be more robust with respect to sparse and noisy data (Song et al., 2023), and where the model is treated as three components: a single feature extractor and two contextualizing networks (student and teacher, Figure 3). Distillation, in general, is the notion of transferring knowledge from a teacher to a student model, whereas mean teacher self-distillation is to update only the student model via gradient descent and let the teacher model track the student's weights using an exponentially moving average (EMA); see Supporting Information: Section 'Mean-teacher distillation' for an intuition on how this works. The feature extractor is domain-specific, and the two contextualizing networks are domain-agnostic transformer architectures (Vaswani et al., 2017). The feature extractor receives the batch of input samples and produces a fixed-size initial representation that is fed to the two contextualizing networks. The teacher receives the full initial representation from the feature extractor, and the student receives the unmasked timesteps from a masked embedding (Figure 3 and Supporting Information: Section 'Mean-teacher distillation'). The teacher produces a target embedding, and the student produces a prediction embedding. The loss function is then a mean-squared-error regression to match the prediction and the target.

For additional information about the transformer architecture, our domain-specific regularization and masking techniques during pretraining, and the pretraining hyperparameters for each setting see Supporting Information: Sections 'Additional details on pretraining and the animal2vec architecture and 'A masking strategy for animals', as well as Table 4.

### 2.2.2 | Fine-tuning animal2vec

For fine-tuning, we largely follow the approach in (Baevski et al., 2020, 2022, 2023), but average the embeddings from all transformer layers rather than just using the output of the last layer, use the focal criterion (Lin et al., 2017) as opposed to cross entropy as the loss function, and use between-classes-Learning (Tokozume et al., 2018) (BCL). See Supporting Information: Section 'Finetuning animal2vec' for further details.

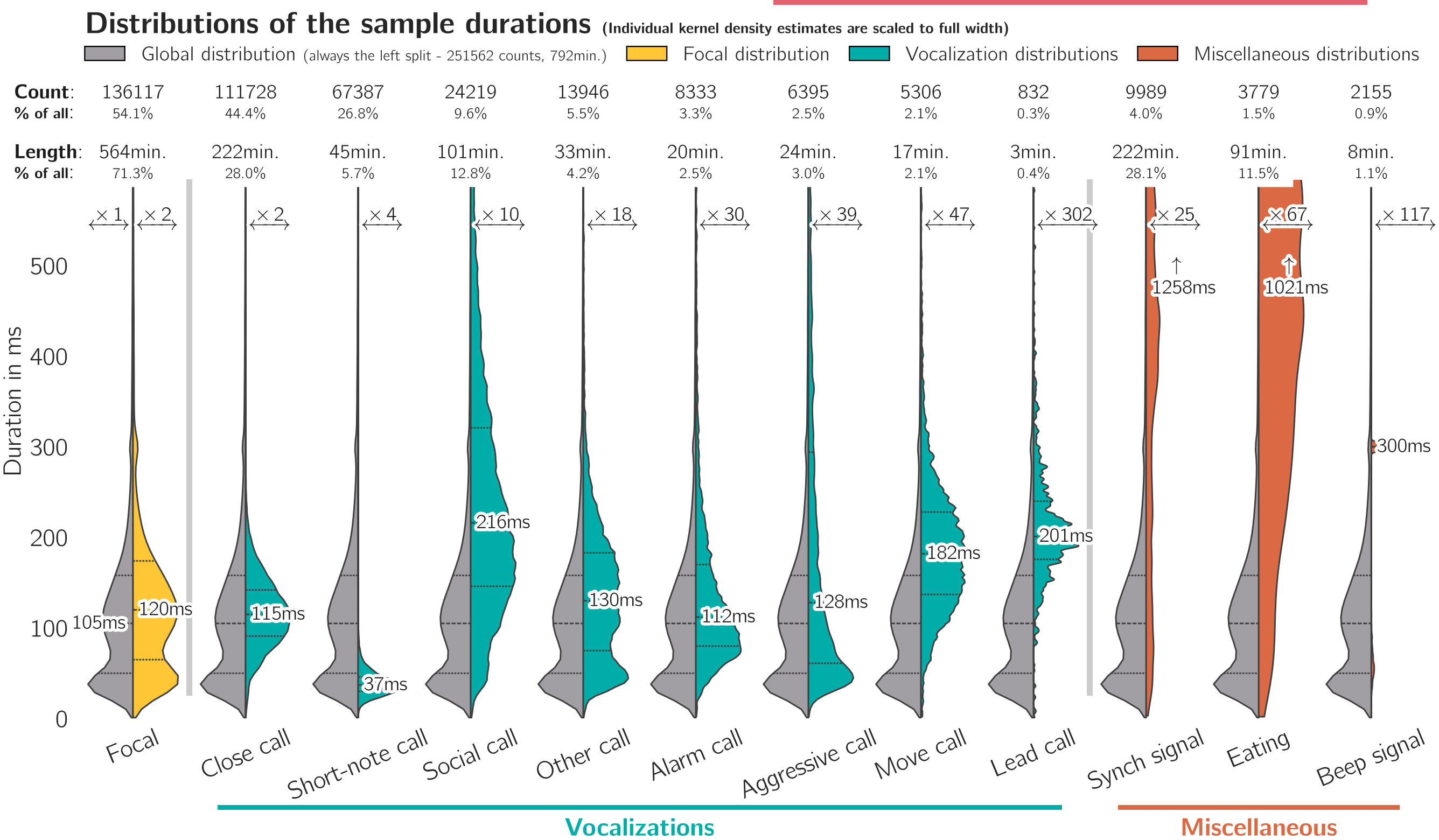

FIGURE 2 The statistics of the MeerKAT dataset. Shown are the temporal distributions of all MeerKAT classes in 12 violin plots. Each category shows kernel density estimates of duration for the class (coloured splits on the right). The global distribution across all categories is shown in grey on the left of each plot to make clear how the label durations of each category relate to the dataset overall. All splits are scaled to full width, where the scaling multiplier is shown at the top of each split, as the number of examples for each category varies considerably. In each split, dashed lines show the 25th, 50th and 75th percentiles, where the 50th percentile (median) value is written next to its dashed line. In addition, the event count, the total duration in minutes and the percentage with respect to all counts/total duration are displayed at the top of each plot.

### 2.2.3 | Evaluating animal2vec

We evaluated and analysed animal2vec on the MeerKAT corpus and the NIPS4BPlus birdsong benchmark (Morfi et al., 2018), and provide a comprehensive overview in Table 1. Scenarios 1 to 6 are performed using MeerKAT, and 7 to 8 use NIPS4BPlus. Scenario 1 is the most extensive test and utilizes stratified fivefold multi-label cross-validation (Sechidis et al., 2011). Final results were averaged and reported with their standard deviation.

We evaluate model performance using precision and recall. Precision quantifies the accuracy of positive predictions (quality), while recall measures the ability to identify all true instances in the data (completeness). These metrics typically exist in a trade-off, as improving one often degrades the other. This relationship is visualized using precision-recall (PR) curves, which plot precision against recall across various model likelihood thresholds (*confidence* of the model). To summarize class-wise performance with a single metric, we compute the average precision (AP) score (Zhu, 2004), a robust estimator for the area under the PR curve (Boyd et al., 2013). The specifics of our metric calculations are detailed in Supporting Information: Section 'Calculating event-based metrics'

With MeerKAT, we derived six scenarios.

1. For class-wise onset/offset prediction, we compared animal2vec directly to the data2vec 2.0 model (Baevski et al., 2023). This baseline, while optimized for human speech on corpora like Librispeech (Panayotov et al., 2015), is architecturally comparable and uses the pretraining scheme we adapted. Both models were pretrained on the full 1068 h of MeerKAT. To assess few-shot performance, we fine-tuned three animal2vec variants using 1%, 25%, and 100% of the labelled subset and compared them against the fully trained data2vec 2.0 model.
2. For the specific task of detecting vocalizations from a focal animal, we benchmarked animal2vec against WhisperSeq (Gu et al., 2024) and data2vec 2.0 (Baevski et al., 2023). WhisperSeq is an event detecting bioacoustic adaptation of the much larger Whisper transformer model (Radford et al., 2022) (769 M parameter compared to 315 M in animal2vec), which was pretrained on 680,000 h of multilingual human speech. For our comparison, we fine-tune the fully pretrained medium Whisper model using the same train/eval splits of MeerKAT as used in scenario 1. However, we use only the *focal* class as binary target label. This creates an event detection experiment where the goal is to determine the focal animal solely from all recorded vocalizations. We compare WhisperSeq against the predictions of animal2vec and data2vec 2.0 for the *focal* class from scenario 1.

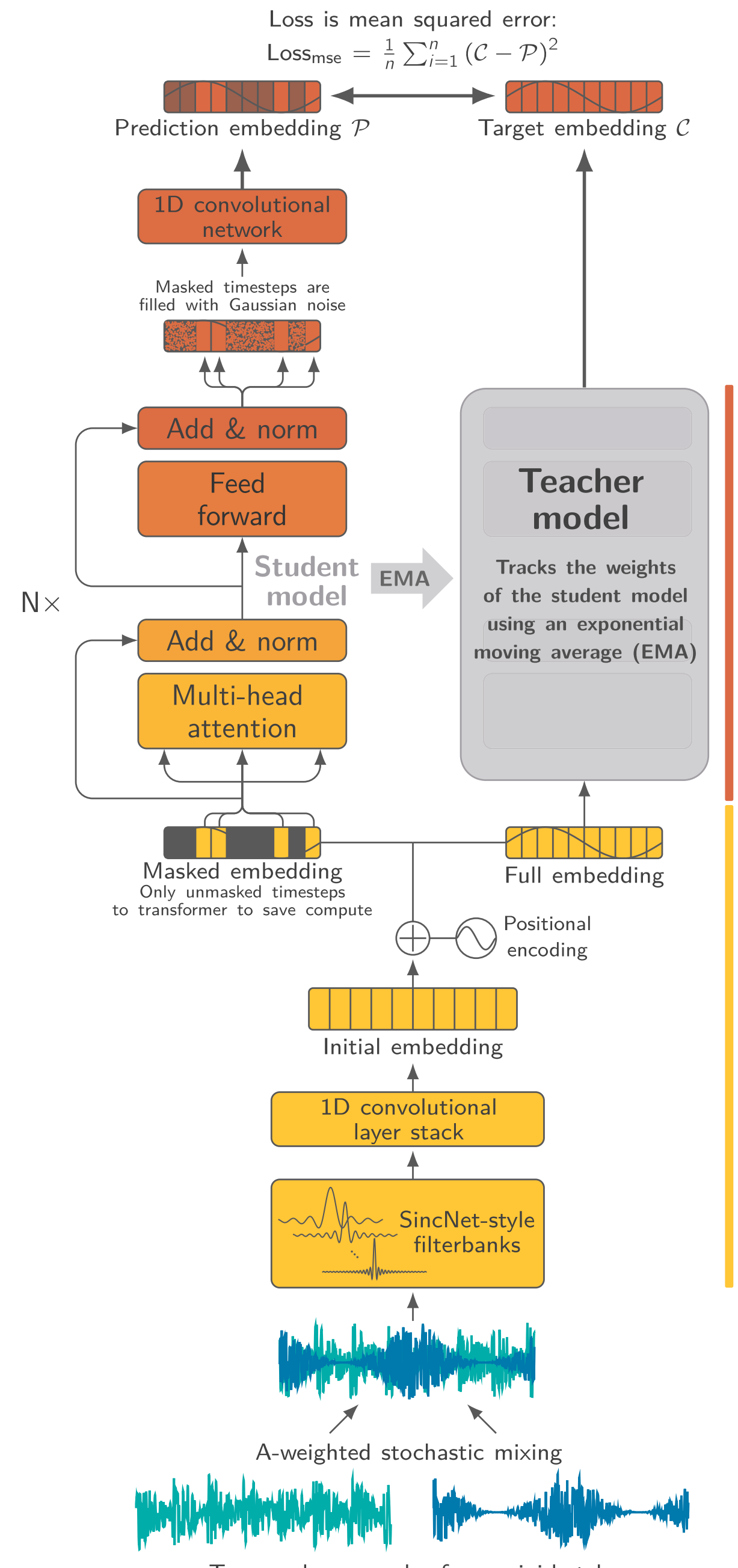

FIGURE 3 The animal2vec self-distillation pretraining scheme. A feature extractor generates initial representations from an A-weighted mixture of two samples from the batch, which are fed to two networks. The teacher network processes the complete, unmasked representation to create target embeddings. The student network receives a masked version and is trained to predict the teacher's targets via a regression loss.

3. A generalizability study was performed by evaluating data2vec 2.0 (Baevski et al., 2023) and animal2vec on a held-out set of data (≈20% of the MeerKAT labelled data) that was not used in either pretraining or fine-tuning, with results presented in the Supporting Information: Section 'Generalization study'.
4. To evaluate the quality of animal2vec's learned representations directly, we conducted a linear evaluation (Chen, Fan, Girshick, & He, 2020; Chen, Kornblith, Swersky, et al., 2020; Oord et al., 2018). We froze the pretrained and fine-tuned animal2vec models and trained a linear support-vector-classifier (SVC) on their output embeddings to assess the linear separability of their feature space (Table S2).
5. An ablation study determined the performance contribution of each of animal2vec's key components, including masking and regularization adaptations, BCL augmentation, layer averaging, and the use of focal loss (Supporting Information: Section 'Ablation study')
6. We demonstrate the temporal and spectral interpretability of the learned parameters of animal2vec by calculating the cumulative frequency response (CFR) of the learned sinc filters (as per (Ravanelli & Bengio, 2018)) and analyse the learned attention maps of animal2vec in Supporting Information: Section 'Temporal and spectral interpretation'.

To assess the transfer learning capabilities of our framework, we also performed two separate experiments on the NIPS4BPlus birdsong benchmark (Morfi et al., 2018) by pretraining animal2vec on a 700 h subset of the xeno-canto dataset (Xeno-canto foundation (officially: Stichting Xeno-canto voor natuurgeluiden), n.d.) and fine-tuning on NIPS4BPlus. Using the numbering scheme of Table 1:

7. We compare animal2vec's class-wise event-based predictions to the pre-segmented sequence classification results of Bravo Sanchez et al. (2021).
8. We calculate binary frame-level predictions from animal2vec's framewise output and compare them against results from Morfi and Stowell (2018).

Results for scenarios 1, 2, 7, and 8 are provided in the Section 3, each within its own subsection. Due to space constraints in the main text, scenarios 3 to 7 are shown in a single *Additional experiments* section in the Supporting Information. We summarize and discuss all results in the Section 4.

## 3 | RESULTS

### 3.1 | Predicting onset/offset in all classes

Overall, animal2vec outperforms data2vec 2.0 even in the 1% few-shot experiment (Figure 4, Table 2a). data2vec 2.0 achieves a precision ≥0.5 for recall values below 0.2 (microaverage AP

TABLE 1 Overview of all experiments and analysis with all datasets and models, and where to find the results in the text.

| # | Task/analysis | Models | Figure/Table |
|---|---|---|---|
| **MeerKAT** | | | |
| 1 | Call type classification and few-shot performance | a2v, d2v2 | Figure 4; Table 2; Figures S6–S17 |
| 2 | Focal animal detection | a2v, d2v2, WS | Table 3 |
| 3 | Generalization study | a2v, d2v2 | Table S1 |
| 4 | Linear evaluation of frozen embeddings | a2v | Table S2 |
| 5 | Ablation study on changes over d2v2 | a2v | Table S3 |
| 6 | Temporal & spectral interpretation | a2v | Figures S1–S3 |
| **NIPS4Bplus** | | | |
| 7 | Predicting pre-segmented sequences | a2v, DN121, RN50, SN, VGG, WCNN | Table 4 |
| 8 | Framewise event prediction | a2v, WHEN | Table 4 |

*Note*: References prefixed with an *S* refer to figures/tables in the supplemental material, where Supporting Information: Section 'Additional experiments' describes all additional experiments. The acronyms for the tested models are: A2v = animal2vec, d2v2 = data2vec 2.0 (Baevski et al., 2023), WS = WhisperSeq (Gu et al., 2024), DN121 = Densenet121 (Huang et al., 2017), RN50 = Resnet50 (He et al., 2016), SN = SincNet (Ravanelli & Bengio, 2018), VGG = VGG16 (Simonyan & Zisserman, 2014), WCNN = Waveform + CNN (Bravo Sanchez et al., 2021) and WHEN = WHEN (MMM) (Morfi & Stowell, 2018).

score is 0.30), whereas even the animal2vec 1% few-shot model never falls below a precision of almost 0.7 (at recall around 0.8; overall AP score is 0.83). The models trained using the 25% and 100% fine-tune splits outperform data2vec 2.0 and the 1% fine-tuning result by a wide margin (overall AP scores of 0.88 and 0.91).

To demonstrate the variation in performance across different event types, we present three representative event classes in the main text. First, the close call class: the easiest to classify as it is very common (most abundant single vocalization) and is comparably long (median duration of 115 ms). Second, the short-note call class: the hardest to classify in terms of duration as it has the shortest median duration (37 ms), yet it occurs frequently (second most abundant vocalization). Third, the alarm call class: the hardest to classify in terms of occurrence (fourth rarest vocalization) but has a comparable median duration to the close call class (112 ms). We present the precision-recall curves of all other classes in Figures S6–S17.

The results for the three selected classes in Figure 4 show animal2vec outperforms data2vec 2.0, achieving AP scores of 0.90, 0.88, and 0.57 using 1% of the data and 0.94, 0.92, and 0.80 using 100% of the data. data2vec 2.0 performs reasonably on the close call class (AP score of 0.49) but achieves low scores with the short-note and alarm call classes (AP scores of 0.14 and 0.03).

As observed in the global average, results using the model trained with 25% and 100% of the available labels are comparable, indicating a saturation effect, where further improvements are not attainable through more labeling. We observe this saturation effect for all but four classes (other, alarm, aggressive, and lead calls). These four classes are among the five rarest classes (Table 2a). The move call class is the only rare class where animal2vec was able to achieve a comparable result in the 25% and 100% setting. We attribute this to the move call class being somewhat easier to predict, having a longer median duration of 182 ms (the third longest of all vocalization classes), but being not so rare as the comparably long lead call class (Figure 2).

The results for the miscellaneous classes (synch, beep, and eating) are comparable to the vocalization classes, except for the synch signal class. There, data2vec 2.0 performs almost on par with animal2vec, achieving an AP score of 0.91. This is expected as a synch signal contains synthetic speech stating the current time. data2vec 2.0 performs well on this class since it was designed for human speech (Baevski et al., 2023). However, even in this case, animal2vec (1%) matches the data2vec 2.0 performance (AP score of 0.89), learning from only 80 labelled examples, compared to the 7990 labelled examples used in data2vec 2.0.

## 3.2 | Detecting focal vocalizations

Results on the focal detection task are given in Table 3, where animal2vec outperforms data2vec 2.0 (Baevski et al., 2023) and the WhisperSeq model (Gu et al., 2024). The latter two reach an F1 score of between 0.56 and 0.57, where WhisperSeq has higher recall and data2vec 2.0 higher precision scores.

## 3.3 | Xeno-canto pretraining and NIPS4Bplus fine-tuning

To demonstrate transfer learning capabilities and assess performance on small-scale datasets, which is challenging for transformer

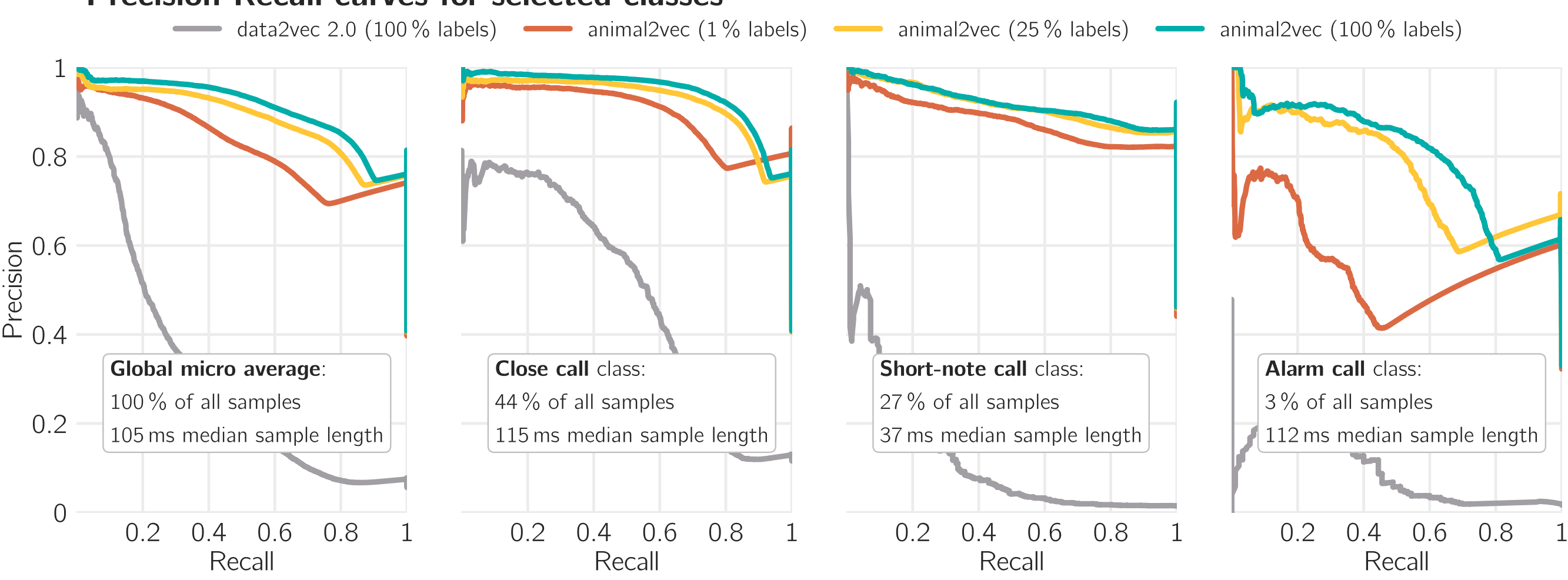

FIGURE 4 The precision-recall curves of animal2vec for selected classes. Shown are four precision-recall curves for (i) the global micro average, (ii) the close call, (iii) the short-note call, and (iv) the alarm call class. Results of animal2vec using 1%, 25%, and 100% of the training data are in red, yellow, and teal, respectively, and data2vec 2.0 results are in grey. Overlays within each subplot show statistics about the occurrence-wise percentage share and the median duration of all events in this class.

TABLE 2 Class-wise dataset statistics and results.

| | (a) Average precision scores [89] | | | | (b) Sample sizes | | | |
|---|---|---|---|---|---|---|---|---|
| | data2vec 2.0 [61] | animal2vec | | | Evaluation | Training | | |
| % Training labels | 100% | 1% | 25% | 100% | – | 1% | 25% | 100% |
| Focal | $0.59_{(2)}$ | $0.86_{(1)}$ | $0.92_{(1)}$ | **0.94**$_{(1)}$ | $24{,}594_{(280)}$ | $983_{(28)}$ | $24{,}650_{(135)}$ | $98{,}520_{(280)}$ |
| *Vocalizations* | | | | | | | | |
| Close call | $0.49_{(1)}$ | $0.90_{(2)}$ | $0.93_{(2)}$ | **0.94**$_{(1)}$ | $22{,}418_{(153)}$ | $907_{(41)}$ | $22{,}342_{(137)}$ | $89{,}310_{(153)}$ |
| Short-note call | $0.14_{(1)}$ | $0.88_{(1)}$ | $0.91_{(1)}$ | **0.92**$_{(1)}$ | $13{,}336_{(158)}$ | $522_{(40)}$ | $13{,}505_{(139)}$ | $54{,}051_{(158)}$ |
| Social call | $0.30_{(2)}$ | $0.65_{(1)}$ | $0.79_{(1)}$ | **0.84**$_{(1)}$ | $4788_{(81)}$ | $207_{(14)}$ | $4847_{(42)}$ | $19{,}431_{(81)}$ |
| Other call | $0.07_{(2)}$ | $0.33_{(3)}$ | $0.43_{(2)}$ | **0.50**$_{(3)}$ | $2754_{(67)}$ | $114_{(18)}$ | $2799_{(58)}$ | $11{,}192_{(67)}$ |
| Alarm call | $0.03_{(1)}$ | $0.57_{(1)}$ | $0.73_{(1)}$ | **0.80**$_{(1)}$ | $1649_{(118)}$ | $71_{(12)}$ | $1704_{(74)}$ | $6684_{(118)}$ |
| Aggressive call | $0.09_{(1)}$ | $0.54_{(2)}$ | $0.62_{(1)}$ | **0.71**$_{(2)}$ | $1214_{(58)}$ | $50_{(15)}$ | $1338_{(28)}$ | $5181_{(58)}$ |
| Move call | $0.09_{(2)}$ | $0.53_{(1)}$ | $0.59_{(2)}$ | **0.61**$_{(1)}$ | $1080_{(23)}$ | $42_{(6)}$ | $1071_{(34)}$ | $4226_{(23)}$ |
| Lead call | $0.01_{(1)}$ | $0.41_{(1)}$ | $0.39_{(2)}$ | **0.50**$_{(1)}$ | $165_{(10)}$ | $8_{(1)}$ | $174_{(12)}$ | $667_{(10)}$ |
| *Miscellaneous* | | | | | | | | |
| Synch signal | $0.91_{(1)}$ | $0.89_{(1)}$ | $0.96_{(1)}$ | **0.98**$_{(2)}$ | $1999_{(2)}$ | $80_{(0)}$ | $1997_{(1)}$ | $7990_{(2)}$ |
| Eating | $0.12_{(1)}$ | $0.59_{(1)}$ | $0.83_{(2)}$ | **0.87**$_{(1)}$ | $760_{(10)}$ | $31_{(2)}$ | $754_{(11)}$ | $3019_{(10)}$ |
| Beep signal | $0.26_{(1)}$ | $0.74_{(1)}$ | $0.77_{(2)}$ | **0.80**$_{(1)}$ | $430_{(5)}$ | $18_{(1)}$ | $431_{(4)}$ | $1725_{(5)}$ |
| Macroaverage | $0.26_{(1)}$ | $0.66_{(1)}$ | $0.74_{(1)}$ | **0.78**$_{(1)}$ | | | | |
| Microaverage | $0.30_{(1)}$ | $0.83_{(1)}$ | $0.88_{(1)}$ | **0.91**$_{(1)}$ | | | | |

*Note*: (a) shows the average precision scores (AP) (Zhu, 2004) of each model, where the given percentage indicates the training sample size during fine-tuning. data2vec 2.0 uses 100% of the training samples for fine-tuning. The strongest result per class is in bold letters. The two bottom rows show the micro- and macroaverage across all classes except *Focal*. (b) It shows the training and evaluation split sample sizes used for fine-tuning. The standard deviation (SD) across the stratified multi-label fivefold cross validation routine (Sechidis et al., 2011) is given in smaller brackets next to each value, where, in (a), the SD refers to the AP scores across the validation splits, and, in (b), the SD refers to the sample number for each class.

architectures that lack inductive bias (Lee et al., 2022), we evaluated animal2vec on the public NIPS4Bplus birdsong dataset. The model was first pretrained on a 700h subset of the xeno-canto database (Xeno-canto foundation (officially: Stichting Xeno-canto voor natuurgeluiden), n.d.), with no overlap with the fine-tuning data (Howard et al., 2020; Xeno-Canto Bird Recordings Extended (A-M), n.d.; Xeno-Canto Bird Recordings Extended (N-Z), n.d.). The NIPS4Bplus dataset contains approximately 1h of densely annotated, multi-label audio across 687 recordings, encompassing 51 bird species and 81 classes. A full description of the dataset is available in (Morfi et al., 2019), and all fine-tuning and pretraining hyperparameters are detailed in Table 4 and Table S5.

We compare our event-based prediction performance (precision, recall, and *F*1 score) against results from previous studies that used different evaluation schemes (Bravo Sanchez et al., 2021; Morfi & Stowell, 2018), such as pre-segmented sequence classification and binary frame-level prediction (Table 4). To facilitate comparison with frame-level methods, we treat any class prediction at a given timestep as a positive event, an approach that likely reduces precision due to an increase in false positives.

animal2vec establishes a new state-of-the-art on the NIPS4Bplus dataset, outperforming the previous best models—Densenet121 (Huang et al., 2017) and WHEN (MMM) (Morfi & Stowell, 2018)—with an increase in *F*1 score of 0.06 and 0.08, respectively.

## 4 | DISCUSSION

In this work, we present animal2vec and MeerKAT and make them openly available. animal2vec is a self-supervised framework and transformer-based model tailored for bioacoustics, while MeerKAT is the largest public dataset on a non-human terrestrial mammal and is specifically designed for the pretrain/fine-tune training paradigm. MeerKAT consists of over 1000 h of audio, of which 184 h have detailed labels, enabling the analysis of event detection performance and noise resilience.

Bioacoustics is, in some respects, a more demanding field than human speech recognition research due to its lack of labelled, large-scale datasets and domain-specific pretraining methods, combined with a focus on extremely rare and brief events of interest. animal2vec attempts to address these challenges and outperforms the data2vec 2.0 (Baevski et al., 2023) and WhisperSeq (Gu et al., 2024) transformer models, both originally devised for human speech, by a large margin on the MeerKAT dataset. WhisperSeq achieved state-of-the-art performance on various bioacoustics datasets (Gu et al., 2024), has more than double the size of animal2vec (769 M parameters vs 315 M), and was pretrained on a far larger pretraining corpus (680,000 h of human speech vs. 1068 h of the MeerKAT dataset in data2vec 2.0 and animal2vec). However, in our tests, the performance gap between animal2vec and WhisperSeq was considerable, suggesting that architectures designed for human speech struggle to compensate for the extreme levels of sparsity and noise found in the MeerKAT corpus, which was our initial motivation for designing the animal2vec framework.

TABLE 3 Results on focal detection task.

| | Performance on focal detection | | |
|---|---|---|---|
| | Precision | Recall | *F*1 |
| data2vec 2.0 [61] | 0.515 | 0.604 | 0.556 |
| WhisperSeq (Gu et al., 2024) | 0.469 | 0.739 | 0.573 |
| animal2vec | **0.905** | **0.801** | **0.849** |

*Note*: Precision and Recall scores are given along with their harmonic mean; the *F*1 score (Van Rijsbergen, 1979). We compare animal2vec against data2vec 2.0 (Baevski et al., 2023) and the WhisperSeq model (Gu et al., 2024). Precision and recall for animal2vec and data2vec 2.0 are taken from Figure S7 of the Supporting Information, where we chose the precision and recall values that maximize the *F*1 score. The strongest result per class is in bold letters.

In addition, animal2vec demonstrates strong few-shot learning capabilities, sometimes requiring only 1% of the available labelled data to achieve competitive results, enabling researchers without access to large amounts of labelled data to effectively fine-tune animal2vec.

Further validation is provided through a series of experiments in the supplemental information. A generalizability study on a held-out test set, excluded even from pretraining, confirmed that animal2vec generalizes well to unseen data from the same domain. A linear evaluation of the model's embeddings revealed that the learned feature space is highly structured and linearly separable even before supervised fine-tuning. Finally, the model's decision-making process is interpretable; the cumulative frequency response (CFR) of its learned sinc filters aligns with the known vocal formants of meerkats, and its transformer attention maps show a sophisticated use of past and future temporal context to make predictions.

To enable comparison with existing datasets, we evaluated a xeno-canto pretrained animal2vec on the NIPS4Bplus dataset, setting a new baseline on two distinct tasks. Notably, it also surpasses a SincNet implementation, which uses a comparable audio-analysis frontend. However, due to its size, animal2vec's performance is contingent on pretraining. Smaller models, such as SincNet (2.6 M parameters), remain a viable alternative, offering a balance of interpretability, computational efficiency, and classification performance.

Moreover, we designed animal2vec as a modular framework, where our novel feature extraction module can be used as a frontend for other models, our transformer model can be used with other frontends, and both can be used with other pretraining or fine-tuning approaches on different datasets or jointly trained with MeerKAT.

The immediate future for animal2vec is (i) to incorporate more data from more species (insects, birds, marine and terrestrial animals), recording environments (marine, avian), using a more diverse set of recorders (passive acoustic monitoring, different portable recorders using different microphones, audio from video traps, community science data; Jäckel et al., 2021) where challenges like the large variability in different sampling rates need to be solved, and (ii) to include more data modalities such as accelerometer and GPS data from next-generation biologging tags (Demartsev et al., 2023), where animal2vec needs to be enabled to make use of such auxiliary data streams but not to decrease in performance when they are missing.

Ultimately, our vision for animal2vec and MeerKAT is for them to serve as a stepping stone towards a next-generation reference work, where, in the future, we envision a foundational-level pretrained

**TABLE 4** Microaverage classification results on the NIPS4Bplus dataset (Morfi et al., 2019).

| Models | Precision | Recall | *F1* |
|---|---|---|---|
| (a) Predicting pre-segmented sequences | | | |
| Densenet121 (Huang et al., 2017) | 0.76 | 0.75 | 0.76 |
| Resnet50 (He et al., 2016) | 0.76 | 0.74 | 0.75 |
| SincNet (Ravanelli & Bengio, 2018) | 0.75 | 0.73 | 0.74 |
| VGG16 (Simonyan & Zisserman, 2014) | 0.74 | 0.73 | 0.74 |
| Waveform + CNN (Bravo Sanchez et al., 2021) | 0.72 | 0.71 | 0.71 |
| animal2vec | **0.81** | **0.88** | **0.84** |
| (b) Framewise event prediction | | | |
| WHEN (MMM) (Morfi & Stowell, 2018) | – | – | 0.74 |
| animal2vec | 0.79 | 0.86 | **0.82** |

*Note*: The metrics for the models trained on pre-segmented sequences are taken from (Bravo Sanchez et al., 2021) and the one for binary timestep prediction is the best result from (Morfi & Stowell, 2018), called *WHEN model using MMM loss*. In addition, we provide results for animal2vec's Onset/Offset/Overlap predictions, using the same methodology described in Supporting Information: Section S3. The strongest result per class is in bold letters.

animal2vec model that researchers can directly use for fine-tuning on their data without the need for large-scale GPU facilities.

While much work remains to achieve this vision, we hope that providing open, accessible, and portable data and code will help stimulate the bioacoustics community to work with us towards achieving this goal.

## AUTHOR CONTRIBUTIONS

The field studies at the Kalahari Research Centre in 2017 and 2019 were performed by Vlad Demartsev, Baptiste Averly, Gabriella Gall, Lily Johnson-Ulrich, Marta B. Manser, and Ariana Strandburg-Peshkin. Data labeling was performed by many student research assistants (see Acknowledgments) under the supervision of Vlad Demartsev, Baptiste Averly, and Ariana Strandburg-Peshkin. Data cleaning and post-processing were performed by Vlad Demartsev and Julian C. Schäfer-Zimmermann. Early studies using different deep learning architectures in combination with the *MeerKAT* dataset were performed by Kiran L. Dhanjal-Adams, Mathieu Duteil, Marie A. Roch, Ariana Strandburg-Peshkin, and Dan Stowell. The conceptual idea for *animal2vec* was conceived by Julian C. Schäfer-Zimmermann, where Marie A. Roch, Ariana Strandburg-Peshkin, and Julian C. Schäfer-Zimmermann jointly analysed the results of *animal2vec*. Julian C. Schäfer-Zimmermann wrote the codebase for *animal2vec*, with guidance from Marie A. Roch, where Marius Faiß and Marie A. Roch conducted a code review before publication. The figures were created by Julian C. Schäfer-Zimmermann. The initial manuscript draft was written by Julian C. Schäfer-Zimmermann. All authors provided edits and feedback on the final manuscript.

## ACKNOWLEDGEMENTS

We thank Rebecca Schäfer for their immense help during the field data collection. We are indebted to many student research assistants for their help with processing and labelling of the audio data (L. Leonardos, C. Maier, J. Denger, L. Batke, S. Eleonori, F. Raabe, H. Brønnvik, J. Ruff, B. Ehrmann, and S. Knab). We are grateful to the Kalahari Research Trust and Northern Cape Department of Environment and Nature Conservation for research permission at the Kalahari Research Centre, as well as the support of the Universities of Cambridge and Zurich and MAVA Foundation on the maintenance of the habituated meerkat population. We thank T. Vink and W. Jubber for organizing the field site and the managers and volunteers of the Kalahari meerkat Project for maintaining habituation and long-term data collection of the meerkats. Open Access funding enabled and organized by Projekt DEAL.

## FUNDING INFORMATION

Max Planck Society; Alexander von Humboldt-Stiftung; Centre for the Advanced Study of Collective Behaviour: EXC 2117–422037984; Human Frontier Science Program: RGP0051/2019; Minerva Foundation; Gips-Schüle Foundation; Young Scholars Fund at the University of Konstanz; Alexander von Humboldt Foundation post-doctoral fellowships; EU MSCA Doctoral Network *Bioacoustic AI* (BioacAI, 101071532); the long-term research on meerkats is currently supported by funding from the European Research Council (ERC) under the European Union's Horizon 2020 research and innovation program (Nos. 742808 and 294494) and a Grant from the Natural Environment Research Council (Grant NE/G006822/1) to Tim Clutton-Brock, as well as by Grants from the University of Zurich to Marta B. Manser and the MAVA Foundation.

## CONFLICT OF INTEREST STATEMENT

The authors declare no competing financial or non-financial interests.

## PEER REVIEW

The peer review history for this article is available at https://www.webofscience.com/api/gateway/wos/peer-review/10.1111/2041-210x.70218.

## DATA AVAILABILITY STATEMENT

The MeerKAT dataset is openly available via the Max-Planck data repository Edmond (Schäfer-Zimmermann et al., 2024; https://doi.org/10.17617/3.0J0DYB) using a CC BY-NC licence (Creative Commons Attribution-NonCommercial 4.0 International). The weights for our pretrained and fine-tuned transformer models are also available via Edmond (https://doi.org/10.17617/3.ETPUKU) using a CC BY-NC licence (Creative Commons Attribution-NonCommercial 4.0 International). The code for animal2vec is available via the official GitHub repository (Official GitHub Repository for animal2vec, n.d.; https://github.com/livingingroups/animal2vec) under an MIT licence (The MIT License) and archived at Zenodo: https://doi.org/10.5281/zenodo.17640321 (Schäfer-Zimmermann, 2025).

## ETHICS AND PERMITS

All procedures were approved by the ethical committees of the University of Pretoria, South Africa (permit: EC031-17) and the Northern Cape Department of Environment and Nature Conservation (permit: FAUNA 1020/2016).

## ORCID

*Julian C. Schäfer-Zimmermann* https://orcid.org/0000-0002-4784-1013

## SUPPORTING INFORMATION

Additional supporting information can be found online in the Supporting Information section at the end of this article.

**Supporting Information S1.** Supplemental information featuring an introduction for people lacking a machine-learning background, the results for the studies 3 to 6 in Table 1 of the main text, and additional Materials and Methods.